\documentclass[review]{elsarticle}

\usepackage{lineno,hyperref}
\usepackage{multirow}
\usepackage{subfigure}
\usepackage{tabularx}
\usepackage{caption} 

\journal{Nucl. Instr. and Meth. A}

\begin{document}

\begin{frontmatter}
\title{Direct method for the quantitative analysis of surface contamination on ultra-low background materials from exposure to dust.}

\author[pnnl]{M.L. di Vacri\corref{mycorrespondingauthor}}
\author[pnnl]{I.J. Arnquist} 
\author[snolab,lu]{S. Scorza} 
\author[pnnl]{E.W. Hoppe} 
\author[snolab,lu]{J. Hall}
\address[pnnl]{Pacific Northwest National Laboratory, Richland, WA 99352, USA.}
\address[snolab]{SNOLAB, Lively, ON P3Y 1N2, Canada.}
\address[lu]{Laurentian University, Department of Physics, Sudbury, ON P3E 2C6, Canada.}
\cortext[mycorrespondingauthor]{marialaura.divacri@pnnl.gov}

\begin{abstract}
In this work we present a method for the direct determination of contaminant fallout rates on material surfaces from exposure to dust. Naturally occurring radionuclides $^{40}$K, $^{232}$Th, $^{238}$U and stable Pb were investigated. Until now, background contributions from dust particulate have largely been estimated from fallout models and assumed dust composition. Our method utilizes a variety of low background collection media for exposure in locations of interest, followed by surface leaching and leachate analysis using inductively coupled plasma mass spectrometry (ICP-MS). The method was validated and applied in selected locations at Pacific Northwest National Laboratory (PNNL) and the SNOLAB underground facility.  
A comparison between data obtained from direct ICP-MS measurements and those estimated from current model-based predictors is also performed. Discrepancies of one order of magnitude or higher are observed between estimated and directly measured accumulation rates.  
\end{abstract}

\end{frontmatter}

\section{Introduction}
Dust is composed of fine particles resulting from the grinding and breaking of materials in the local environment. It includes soil and minerals from rocks, small amounts of pollen, human and animal hair and skin cells, fibers and debris from the materials found in the environment. The chemical composition of dust generally reflects the local chemical composition of soil and rocks, but it can vary depending on local anthropogenic activities and the environment (e.g. indoor vs outdoor or ordinary environment vs cleanrooms). Interest in investigating dust composition is generally related to air quality and health risk assessments, especially in indoor and outdoor environments where the concentration of dust is significantly high and/or there is a risk associated with the presence of heavy metals, as reported in~\cite{1}, \cite{2} and \cite{3}. Dust is also a primary concern for a variety of applications. Studies have been performed to investigate and control the effects of dust particulate deposition on artworks, as reported in \cite{4}. The semiconductor industry is also well known to be susceptible to particulate contamination \cite{5}\cite{6}. When there is a need to reduce the impact of particle deposition on critical materials and surfaces, operations are typically conducted in cleanrooms. Though designed to maintain extremely low levels of particulates, cleanroom facilities are not totally free from particulate fallout \cite{7}. While still limited research has been conducted to predict the contribution of particulate contamination in cleanrooms for very critical applications, the work in \cite{8} reports one of the most accepted models for particle fallout predictions in cleanrooms. 

In the realm of rare event physics detectors, for which the purity of all materials included in the detector is a very critical aspect, dust fallout on material surfaces and its contribution to radioactive background is of significant concern. The construction of successful ultralow background (ULB) detectors implies extensive ultrasensitive assay campaigns to select the purest materials meeting the extremely stringent radiopurity levels required for rare event detection~\cite{9}-\cite{13}. The major contributors to material radioactive backgrounds are the naturally-occurring primordial radionuclides $^{40}$K, $^{232}$Th and $^{238}$U and their daughters~\cite{14}. Limits for these elements in ULB materials can be very demanding, in the $\mu$Bq$\cdot$kg$^{-1}$ regime or even lower, with greater importance given to material closer to the active target and the larger the mass incorporated into the detector. Isotope $^{210}$Pb, the relatively long-lived ($\lambda$ of 22 years) progeny in the $^{238}$U decay series, is oftentimes also a concern and can be significantly out of secular equilibrium from $^{238}$U. For example, levels of $^{210}$Pb in modern, refined lead can be significantly higher (six orders of magnitude) than levels assumed from the $^{238}$U decay series \cite{15}\cite{16}. For this reason, and the fact that large amounts of Pb shielding are used in ULB experiments, we chose to monitor stable Pb fallout rates from dust. Much effort is being dedicated to predict and control the background contribution from dust to material contamination in cleanroom facilities~\cite{17}-\cite{19}. Predictions typically rely on previously developed models for dust fallout, such as those described in~\cite{8} and  \cite{17}-\cite{19}, and assumed dust elemental composition and other physical variables. 
This work demonstrates a method for the direct determination of radiocontaminant accumulation rates on material surfaces from dust.  

The method includes the exposure of a material to dust, followed by dissolution of deposited contamination in 5\% nitric acid and solution analysis via ICP-MS. Accumulation rates (e.g., ng·day$^{-1}\cdot$cm$^{-2}$) or the analytes of interest are determined based on the measured quantity of analyte, exposed material surface area and time of exposure. Accumulation rates in terms of radioactivity (e.g., $\mu$Bq$\cdot$day$^{-1}\cdot$cm$^{-2}$) are calculated from the specific activities (Bq·g$^{-1}$) of the analytes. It should be noted that ICP-MS analysis does not detect radioactive decays, but, instead, detects the atoms (more specifically the ions) after being mass resolved from concomitant ions in the sample.  
ICP-MS is a useful technique for the ULB physics community in reaching $\mu$Bq/kg sensitivities for radionuclides with very long half-lives (e.g. $>$10$^{6}$ years, like $^{232}$Th, $^{238}$U, and $^{40}$K). Due to the tiny quantities of isotope $^{210}$Pb, there are just not enough atoms to efficiently detected this radionuclide directly by ICP-MS. In this work, only stable Pb is directly measured via ICP-MS. Accumulation rates of $^{210}$Pb are inferred from measured stable Pb accumulation rates, assuming a specific activity of 200~Bq $^{210}$Pb per kg of modern, refined lead from a previous work \cite{20}. It is worth pointing out that we are only studying the $^{210}$Pb contamination from dust falling out of the air and depositing on material surfaces. This work is not meant to investigate $^{210}$Pb contaminations on surfaces from radon progeny implantation. 

 A variety of locations for studying radiocontaminant accumulation rates from dust were investigated at Pacific Northwest National Laboratory (PNNL, Richland, Washington) and SNOLAB (Ontario, Canada), a deep underground class 2,000 cleanroom, excavated at a depth of 6,800~ft in the working Creighton nickel mine, hosting neutrino and dark matter experiments \cite{32}. This study provides a straightforward, ultrasensitive pathway to assess background contributions from particulate dust fallout. 
 
\section{Experimental}
Ultralow background perfluoroalkoxy alkane (PFA) screw cap vials from Savillex (Eden Prairie, MN) were used as containers for sample preparations, preparation of reagent solutions and as an exposed dust collection media. Square silicon coupons (22~mm per side) were cut from a 100~mm diameter, 500~$\mu$m thick, bare silicon wafer from Virginia Semiconductor (Fredericksburg, VA). The wafer was polished on one side.  The polished side of the coupons was used as the exposed surface. Silicon wafers have been utilized in the past for dust deposition studies in cleanrooms~\cite{5}. They are typically low background materials~\cite{21}. Labware rinsing and preparation of reagent solutions were performed using 18.2~M$\Omega\cdot$cm deionized water from a MilliQ system (Merck Millipore GmbH, Burlington, MA, USA).  Optima grade nitric and hydrochloric acid (Fisher Scientific, Pittsburg, PA, USA) were used. Standard solutions of $^{229}$Th and $^{233}$U (Oak Ridge National Laboratory, Oak Ridge, TN, USA) were used to quantify Th and U using an external calibration curve. Non-natural isotopes were chosen as calibration standards to avoid carry over effects from natural Th and U standard solutions, given the extremely low signals to be detected. Standard solutions of K, Ca, Pb (Inorganic Ventures, Christiansburg, VA, USA) and Fe (High Purity Standards, North Charleston, SC, USA) were used for quantitation of these elements through an external calibration curve. All labware involved in the study – vials, bottles, pipette tips, tongs and materials exposed to dust – underwent cleaning and validation before use. A preliminary cleaning was performed in a 2\%v/v Micro90\texttrademark~ detergent (Cole-Parmer, Vernon Hills, IL, USA) solution, followed by multiple rinsing with MilliQ water. Leaching in 3M~HCl and 6M~HNO$_{3}$ solutions preceded a validation to ensure sufficiently low backgrounds. The validation step consisted of pipetting a small volume of 5\%v/v HNO$_{3}$ into each container, closing, shaking and allowing to incubate at 80$^{\circ}$C for at least 12 hours. Tongs, pipette tips and silicon wafer coupons were soaked in 5\%v/v HNO$_{3}$ using validated PFA vials and containers. The leachate was then analyzed via ICP-MS. Any labware failing validation underwent additional cycles of leaching and validation tests until meeting background requirements. Leachates of Si coupons and PFA vials measured prior to surface exposure were used as process blanks. Determinations of K, Ca, Fe, Pb, Th and U were performed using an Agilent 8900 ICP-QQQ-MS (Agilent Technologies, Santa Clara, CA), equipped with an integrated autosampler, a microflow PFA nebulizer and a quartz double pass spray chamber. For lead, thorium and uranium determinations, plasma, ion optics and mass analyzer parameters were optimized based on the instrumental response from a standard tuning solution from Agilent Technologies containing ca.~0.1~ng$\cdot$mL$^{-1}$ of Li, Mg, Co, Y, Ce, Tl. The instrumental response from Tl was used as a reference signal, in order to maximize the signal to noise ratio in the high m/z range. Oxides were monitored and kept below 2\% based on the m/z=156 and m/z=140 ratio (CeO$^+$/Ce$^+$) from Ce in the tuning solution. Potassium, calcium and iron determinations were performed in cool plasma with NH$_{3}$ reaction mode. Instrumental parameters were optimized based on the instrumental response from a solution containing ca.~1~ng$\cdot$g$^{-1}$ K and 0.1~ng$\cdot$g$^{-1}$ Ca and Fe, in-house diluted from a stock solution. Detection limits (DLs) were calculated as 3$\cdot$StdDev of n=3 process blanks (PBs). Detection limits for each analyte are reported in Table~\ref{tab:1}. All sample solutions were measured above detection limits.

\begin{table}
\centering
\begin{tabular}{| c | c |}
\hline 
Analyte & DL \\
\hline
K [fg·g$^{-1}$]  & 30.0 \\
Ca [pg·g$^{-1}$] & 0.82\\  
Fe [pg·g$^{-1}$] & 0.66 \\
Pb [fg·g$^{-1}$] & 70.8 \\
Th [fg·g$^{-1}$] & 0.61 \\
U [fg·g$^{-1}$] & 0.79 \\
\hline
\end{tabular}
\caption{ICP-MS detection limits for K, Ca, Fe, Pb, Th and U, measured as 3$\cdot$StdDev of n=3 PBs, for each analyte.}
\label{tab:1}
\end{table}

Surfaces were exposed over a time period of about a month or longer. After that, vials were filled with 1.5~mL of 5\%v/v HNO3 solution, recapped, shaken and left filled over a few hours for contaminant dissolution. Extreme care was taken during this operation not to disturb any deposition on the surface. No visible dust was noticed on the collected media. Silicon coupons were carefully transferred into validated clean PFA vials. During the transfer, coupons were handled only from the very end edges using validated ultraclean tongs, making sure the exposed surface was facing upward during the transfer. Vials containing the coupons were then filled with 5~mL of 5\%v/v HNO3 solution and capped. The coupons were left fully submerged in the nitric acid solution for a few hours. Vials were often shaken, in order to collect and dissolve all the contamination deposited on the silicon surface.  Solutions were analyzed with ICP-MS for K, Ca, Fe, Pb, Th and U and accumulation rates for these elements were calculated. The total surface from both vials and cap was considered for an exposed surface for one PFA vial sample. For the silicon surface, each coupon represented a sample and only the upward exposed surface area was considered. We assumed no dust deposited on the face of the coupon in contact with the clean wipe and did not include contributions from the negligible side surface of the coupons (500~$\mu$m thickness). 


\section{Results}

\subsection{Exposures at PNNL and SNOLAB }
\begin{figure}[htb!]
    \centering
    \includegraphics[width=0.70\textwidth]{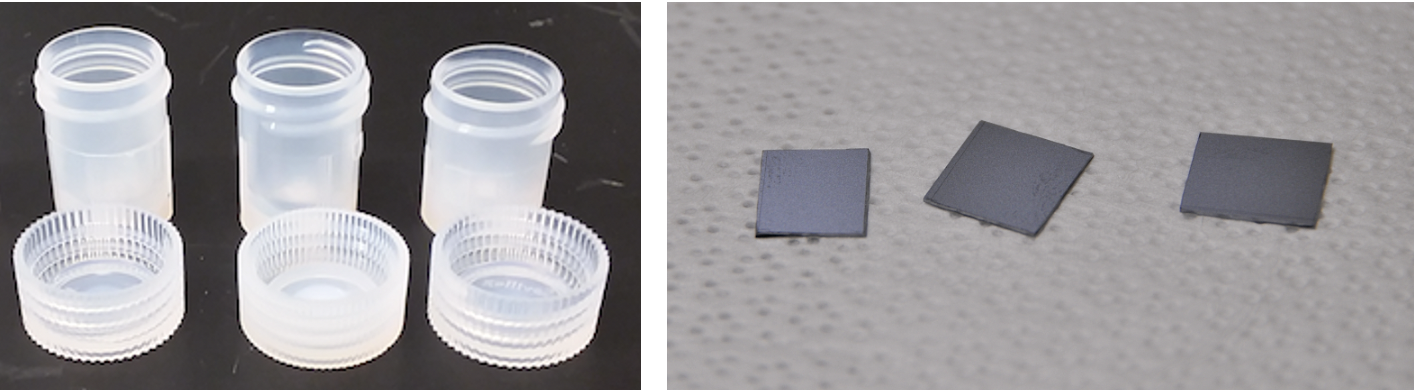}
    \caption{Left: PFA vials exposed to dust in a class 10,000 cleanroom at PNNL. In order to maximize the exposed surface, both vial and cap were facing upward. The total exposed surface for each vial (vial and cap) is ca. 19 cm$^{2}$. Right: silicon coupons exposed to dust in the same class 10,000 cleanroom, in an adjacent location. The total exposed surface for each coupon, only accounting for the upward exposed face, is $ca.$ 5 cm$^{2}$.}
    \label{fig:1}
\end{figure}

The method of dust collection and analysis was preliminarily validated exposing both dust collection media (PFA vials and silicon wafer coupons) to different class cleanroom settings at PNNL, including non-cleanrooms. Validation aimed at assessing the dependence of the method on the material surface, as well as the reliability of obtained results. Sets of three PFA vials and three silicon coupons were exposed to air in a class 10,000 cleanroom at PNNL in adjacent locations, as shown in Figure~\ref{fig:1}. 

\begin{figure}[htb!]
    \centering
    \includegraphics[width=0.65\textwidth]{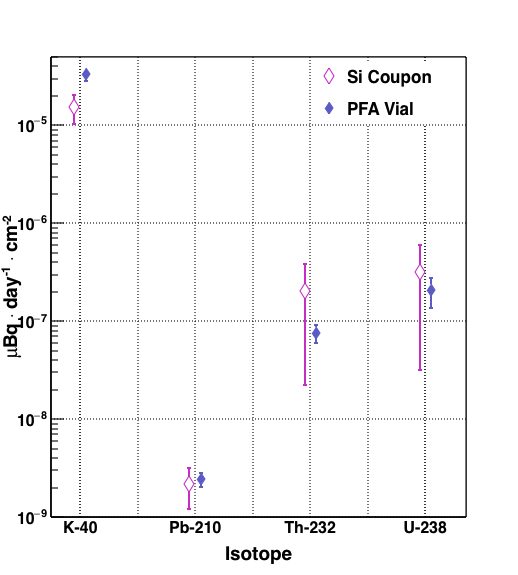}
    \caption{Measured accumulation rates in $\mu$Bq per day per square cm of $^{40}$K, $^{210}$Pb, $^{232}$Th and $^{238}$U on PFA and Si surfaces. Results are reported as the average and standard deviation of three independent replicates. Natural potassium and stable lead were measured with ICP-MS. Accumulation in terms of radioactivity for these elements were calculated based on their specific activities. Activities from $^{210}$Pb were inferred assuming 200~Bq $^{210}$Pb$\cdot$kg$^{-1}$ of modern, refined Pb~\cite{20}.}
    \label{fig:2}
\end{figure}

Results from a 29-day exposure in a class 10,000 cleanroom at PNNL are shown in Figure~\ref{fig:2} and listed in Table~\ref{tab:2}. Vials and coupons were exposed in very adjacent locations, at same height. 
Results are reported, as the average and standard deviation of the three replicates of each set (PFA and Si), in terms of radioactivity accumulation rates in $\mu$Bq$\cdot$day$^{-1}\cdot$cm$^{-2}$. Natural potassium and stable (natural) lead were measured by ICP-MS; their contributions in terms of radioactivity were estimated based on their specific activities. A value of 200~Bq $^{210}$Pb$\cdot$kg$^{-1}$ of modern commercial lead was assumed from a previous study~\cite{20}. 
Results show no significant difference between the accumulation rates on the two types of surfaces for Th, U and Pb. A minor discrepancy (of a factor of $\approx$1.5) was observed for the accumulation rates of potassium on the two material surfaces. Results obtained from the silicon surface exposure show a wider spread compared to those from PFA surface. Silicon coupons only involve the exposure of a horizontal surface, while also vertical (inner) surfaces are involved in the exposure of vials (Figure 1). The good correlation between results obtained from two collection media of different nature and shape demonstrated the independence of the method on the nature of the surface and showed that not only horizontal surfaces are susceptible of dust accumulation.  

\begin{table}
\resizebox{\textwidth}{!}{
\begin{tabular}{|c|c|c|c|c|}
\hline 
 \multirow{2}{*}{Location } &	\multicolumn{4}{|c|}{Accumulation rate [$\mu$Bq$\cdot$day$^{-1}\cdot$cm$^{-2}$]}\\
 & 	$^{40}$K & $^{210}$Pb & $^{232}$Th & $^{238}$U\\
\hline
PFA vial &	(3.3 $\pm$ 0.5)$\cdot$10$^{-5}$ &	(2.4 $\pm$ 0.4)$\cdot$10$^{-9}$&	(7.6 $\pm$ 1.5)$\cdot$10$^{-8}$&	(2.1 $\pm$ 0.7)$\cdot$10$^{-7}$\\
Si Coupon &	(1.5 $\pm$ 0.5)$\cdot$10$^{-5}$ &	(2.1 $\pm$ 0.9)$\cdot$10$^{-9}$&	(2.0 $\pm$ 1.8)$\cdot$10$^{-7}$ &	(3 $\pm$ 3)$\cdot$10$^{-7}$\\
\hline
\end{tabular}}
\caption{Accumulation rates values, in $\mu$Bq per day per square cm, measured for $^{40}$K, $^{210}$Pb, $^{232}$Th and $^{238}$U on PFA and Si surfaces. Results are reported as the average and standard deviation of three independent replicates. Natural potassium and stable lead were measured with ICP-MS. Accumulation in terms of radioactivity for these elements were calculated based on their specific activities. A value of 200~Bq $^{210}$Pb$\cdot$kg$^{-1}$ Pb was considered for Pb~\cite{20}.}
\label{tab:2}
\end{table}

Results were compared to data reported by the U.S. Geological Survey (USGS)~\cite{22} for the relative content of K, Th and U in the soil of the considered area (Pacific Northwest, PNW). According to the USGS, potassium content in the PNW area (level of 10$^{4}$ ppm) is $ca.$ four orders of magnitude higher than that of thorium and uranium, whose concentrations are of the same order of magnitude (ppm levels). Data for Pb are not reported. Dust deposited in cleanrooms looks to be slightly higher in K and/or lower in U/Th relative to distributions seen in surface soil in the area.  Potassium contaminations in cleanrooms are more likely introduced by particulate from human skin cells of cleanroom users, fibers from their garb and grinding of polymers and other materials handled in the space (instrumentation, labware, etc.) rather than particulate from soil. The minor discrepancy between accumulation rates of potassium on PFA and Si surfaces could be explained with a different source, and static effect, of particulate introducing K contaminations compared to particulate carrying Th and U. Particle count was performed regularly, twice per week, in the cleanroom during the exposure time. While nominally a class 10,000 cleanroom, results from particle count indicated an average cleanroom class better than 1,000.

\begin{figure}[htb!]
    \centering
    \includegraphics[width=0.65\textwidth]{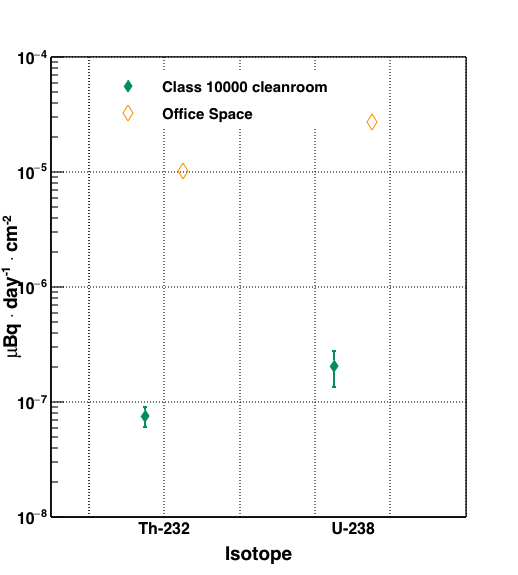}
    \caption{Measured accumulation rates in $\mu$Bq per day per square cm of $^{232}$Th and $^{238}$U on PFA surfaces in an office space and in a class 10,000 cleanroom at PNNL. Results are reported as the average and standard deviation of eight independent replicates.}
    \label{fig:3}
\end{figure}

A set of six PFA vials was also exposed to dust in a class 10 laminar flow hood at PNNL over a 30-day period. Concentrations of K, Pb, Th and U in solutions were measured within instrumental background levels, showing no significant contamination was introduced by dust. Data are therefore not plotted. 

A second test was conducted comparing accumulation rates for $^{232}$Th and $^{238}$U in a class 10,000 cleanroom and in a non-cleanroom environment (e.g., an office space). A set of eight PFA vials was exposed to dust in an office space at PNNL for 30~days and compared to a set of eight PFA vials exposed in a class 10,000 cleanroom at PNNL over the same period. Accumulation rates for $^{232}$Th and $^{238}$U are shown in Figure~\ref{fig:3}. They were measured at (1.0$\pm$0.6)$\cdot$10$^{-8}$ and (2.7$\pm$0.4)$\cdot$10$^{-8}$ $\mu$Bq$\cdot$day$^{-1}\cdot$cm$^{-2}$ for Th and U respectively in the office space. Accumulation rates for $^{232}$Th and $^{238}$U in the class 10,000 cleanroom were (7.6$\pm$1.5)$\cdot$10$^{-11}$ and (2.1$\pm$0.7)$\cdot$10$^{-10}$ $\mu$Bq$\cdot$day$^{-1}\cdot$cm$^{-2}$, respectively. An accumulation rate scaling of a factor of 100 was observed between a class 10,000 cleanroom and an office space at PNNL. Non-cleanroom environments are typically  considered  of class 1,000,000 \cite{7}, although particle concentration highly depends on geographic location, daily weather conditions, ongoing activities and building ventilation.  


Measurements of dust accumulation rates have been performed at the SNOLAB underground research facility, for selected areas. Dust collection was also performed in an office space at the SNOLAB surface building. SNOLAB maintains a series of fixed air particulate meters (MET ONE 6000P) at various places in the laboratory. The data from these sensors are made available to the astroparticle physics community and  SNOLAB users. In addition, a set of twelve witness plates are located around the laboratory at an average height of about 8 feet. Tape lifts are used to measure dust levels on the witness plates with an interval between lifts of about a month~\cite{23}. For the study presented herein, whenever possible, we have investigated the same locations as the fixed dust monitors and witness plates with preference for the most sensitive locations, close to the current and future experiments sites. Figure~\ref{fig:4} shows a map of the SNOLAB underground facility along with the selected locations tagged with a red star and a letter notation. 

\begin{minipage}{0.5\textwidth}
    \includegraphics[width=\textwidth]{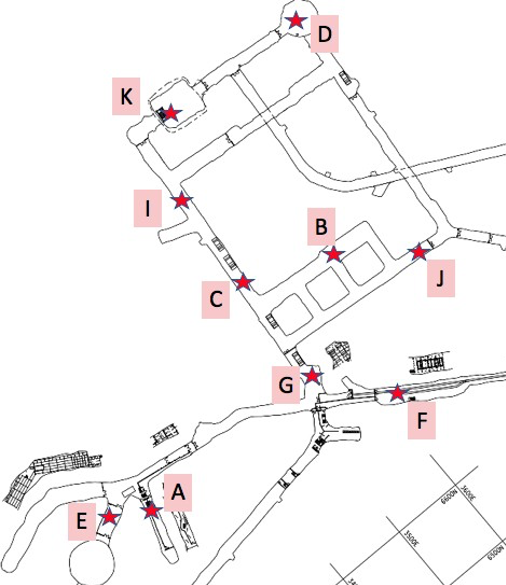}
    \captionof{figure}{Map of the SNOLAB underground facility. Investigated locations are tagged with a red star and a letter notation.}\label{fig:4}
\end{minipage}
\hspace{0.1in}
\begin{minipage}{0.5\textwidth}
\begin{itemize}
\item[]{\textbf{A: South Drift LBL}.} 
\item[]{\textbf{B: Room 127}.}
\item[]{\textbf{C: Drift F}.}  
\item[]{\textbf{D: Room 141}.} 
\item[]{\textbf{E: SNO+ control room}.}
\item[]{\textbf{F: Room 104}.}  
\item[]{\textbf{G: Room 123}.}  
\item[]{\textbf{I: Drift F/J}.}  
\item[]{\textbf{J: Room 132}.}  
\item[]{\textbf{K: Room 137}.}
\end{itemize}
\end{minipage}
\vspace{0.2in}

\begin{figure}[htb!]
    \centering
    \includegraphics[width=0.95\textwidth]{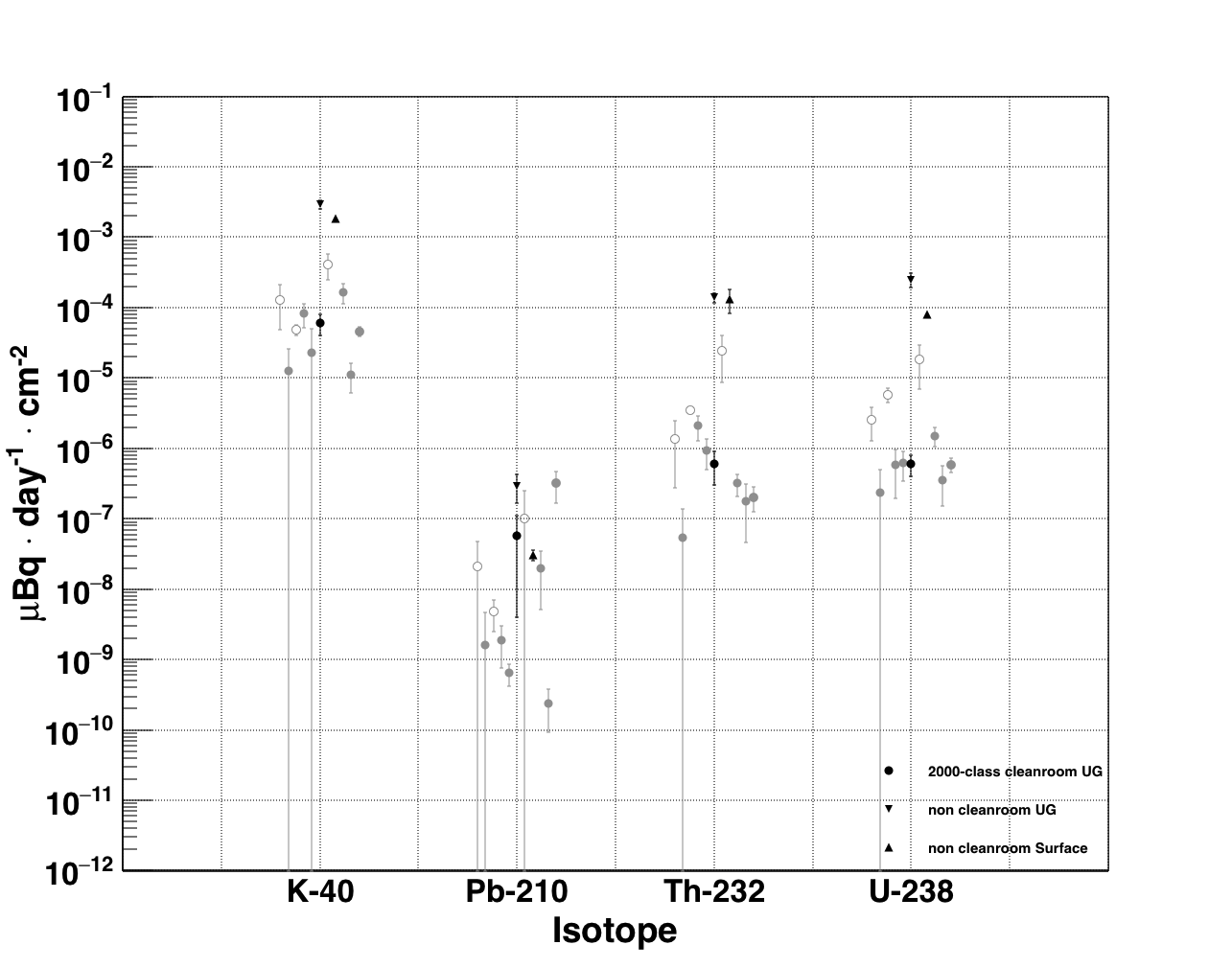}
    \caption{Measured accumulation rates in $\mu$Bq per day per square cm for $^{40}$K, $^{210}$Pb, $^{232}$Th and $^{238}$U $^{238}$U on PFA surfaces in selected locations at the SNOLAB facility. Results are reported as the average and standard deviation of four independent replicates. Activities from $^{210}$Pb were inferred assuming 200~Bq $^{210}$Pb$\cdot$kg$^{-1}$ of modern, refined Pb~\cite{20}.}
    \label{fig:5}
\end{figure}

A brief description of the underground locations is listed below~\cite{24}:  
\begin{itemize}
\item[]{\textbf{A: South Drift LBL}. \\ 
It is a space in the mezzanine level in the south drift, where the low background screening facilities are located (HPGe detectors, XRF, alpha counters).}

\item[]{\textbf{B: Room 127}. \\
Ladder lab, between the CUTE and SuperCDMS SNOLAB experiment locations.}

\item[]{\textbf{C: Drift F}.\\  
Main laboratory hallway, close to PICO-50 experimental area.} 

\item[]{\textbf{D: Room 141}. \\ 
Bottom part of the cryopit which will be hosting the next generation neutrinoless double beta decay experiment.}

\item[]{\textbf{E: SNO+ control room}.} 

\item[]{\textbf{F: Room 104}. \\ 
Transition area (“dirty carwash”) between the mine drift and the class 2,000 cleanroom facility (non-cleanroom environment).} 

\item[]{\textbf{G: Room 123}. \\ 
Junction area right outside the refuge station lunchroom.} 

\item[]{\textbf{I: Drift F/J}.  \\
Main hallway, close to SENSEI and DAMIC experimental areas.}

\item[]{\textbf{J: Room 132}. \\ Chemistry laboratory.} 

\item[]{\textbf{K: Room 137}.\\ Deck on top of the DEAP water tank.} 
\end{itemize}
All locations are class 2,000 cleanrooms, except location F, a transition area from the mine drift to the clean area where the cleaning of materials occurs before access to the laboratory. An office space on the third floor of the surface building (non-cleanroom, labelled as location H in Table \ref{tab:3}) was also investigated. Sets of four PFA vials were used for dust collection in each location and exposed over a 42-day period, except locations D and K, where the exposure lasted 50 days. Exposure was conducted during the 20-day Creighton mine shutdown in August 2018, when very limited activities were being carried on in the laboratory. Measured accumulation rates for $^{40}$K, $^{210}$Pb, $^{232}$Th and $^{238}$U, in $\mu$Bq$\cdot$day$^{-1}\cdot$cm$^{-2}$ are plotted in Figure~\ref{fig:5} and reported in Table~\ref{tab:3}. As stated previously, a specific activity of 200~Bq$\cdot$kg$^{-1}$ was assumed for stable Pb~\cite{20}.
An average of the four replicates measured values along with its standard deviation for each element and location are quoted. In Figure~\ref{fig:5}, the above ground non-cleanroom location is represented with an upward black triangle, while the underground non-cleanroom location is marked with a downward black triangle. Underground class-2,000 cleanroom areas are indicated with circles. Empty grey circle markers refer to underground cleanroom areas where some activities during the exposure time might have triggered higher accumulation rates for $^{40}$K, $^{232}$Th and $^{238}$U (locations B, G and J, later explained) compared to all the other underground class 2,000 cleanroom locations, indicated with a full grey circle marker.  Full black circle markers in Figure~\ref{fig:5} show the average accumulation rates along with their standard error of the mean (SEM) considering the class 2,000 cleanrooms shown as full grey circle markers.  

\begin{table}[h!]
\resizebox{\textwidth}{!}{
\begin{tabular}{|c|c|c|c|c|}
\hline 
 \multirow{2}{*}{Location } &	\multicolumn{4}{|c|}{Accumulation rate [$\mu$Bq$\cdot$day$^{-1}\cdot$cm$^{-2}$]}\\
 & 	$^{40}$K & $^{210}$Pb & $^{232}$Th & $^{238}$U\\
\hline
A& 
(4.6 $\pm$ 0.7)$\cdot10^{-5}$ 
&
(3.2 $\pm$ 1.5)$\cdot10^{-7}$ 
&
(2.0 $\pm$ 0.8)$\cdot10^{-7}$ 
&
(5.9 $\pm$ 1.4)$\cdot10^{-7}$ \\
B 
&
(1.3 $\pm$ 0.8)$\cdot10^{-4}$ 
&
(2 $\pm$ 3)$\cdot10^{-8}$ 
&
(1.3 $\pm$ 1.1)$\cdot10^{-6}$ 
&
(2.6 $\pm$ 1.3)$\cdot10^{-6}$ \\
C 
&
(1.1 $\pm$ 0.5)$\cdot10^{-5}$ 
&
(2.4 $\pm$ 1.4)$\cdot10^{-10}$ 
&
(1.8 $\pm$ 1.3)$\cdot10^{-7}$ 
&
(4 $\pm$ 2)$\cdot10^{-7}$ \\
D 
&
(1.3 $\pm$ 1.4)$\cdot10^{-5}$ 
&
(2 $\pm$ 3)$\cdot10^{-9}$ 
&
(5.5 $\pm$ 0.8)$\cdot10^{-8}$ 
&
(2 $\pm$ 3)$\cdot10^{-7}$ \\
E 
&
(1.6 $\pm$ 0.5)$\cdot10^{-4}$ 
&
(1.9 $\pm$ 1.4)$\cdot10^{-8}$ 
&
(3.2 $\pm$ 1.1)$\cdot10^{-7}$ 
&
(1.5 $\pm$ 0.5)$\cdot10^{-6}$ \\
F 
&
(2.9 $\pm$ 0.4)$\cdot10^{-3}$ 
&
(2.9 $\pm$ 1.3)$\cdot10^{-7}$ 
&
(1.3 $\pm$ 0.2)$\cdot10^{-4}$ 
&
(2.5 $\pm$ 0.6)$\cdot10^{-4}$ \\
G 
&
(4.9 $\pm$ 0.8)$\cdot10^{-5}$ 
&
(5 $\pm$ 2)$\cdot10^{-9}$ 
&
(3.5 $\pm$ 0.2)$\cdot10^{-6}$ 
&
(5.8 $\pm$ 1.3)$\cdot10^{-6}$ \\
H 
&
(1.83 $\pm$ 0.06)$\cdot10^{-3}$ 
&
(3.0 $\pm$ 0.5)$\cdot10^{-8}$ 
&
(1.3 $\pm$ 0.5)$\cdot10^{-4}$ 
&
(8.0 $\pm$ 0.6)$\cdot10^{-5}$\\
I 
&
(8 $\pm$ 3)$\cdot10^{-5}$ 
&
(1.9 $\pm$ 1.1)$\cdot10^{-9}$ 
&
(2.1 $\pm$ 0.8)$\cdot10^{-6}$ 
&
(6 $\pm$ 4)$\cdot10^{-7}$ \\
J 
&
(4.1 $\pm$ 1.6)$\cdot10^{-4}$ 
&
(9.9 $\pm$ 0.1)$\cdot10^{-8}$ 
&
(2.4 $\pm$ 1.5)$\cdot10^{-5}$ 
&
(1.8 $\pm$ 1.1)$\cdot10^{-5}$ \\
K 
&
(2 $\pm$ 3)$\cdot10^{-5}$ 
&
(6 $\pm$ 2)$\cdot10^{-10}$ 
&
(9 $\pm$ 4)$\cdot10^{-7}$ 
&
(6 $\pm$ 3)$\cdot10^{-7}$\\
\hline
\end{tabular}}
\caption{Accumulation rates values, in $\mu$Bq per day per square cm, measured for $^{40}$K, $^{210}$Pb, $^{232}$Th and $^{238}$U on PFA surfaces in selected locations at the SNOLAB facility. Results are reported as the average and standard deviation of four independent replicates. Accumulation in terms of radioactivity for these elements were calculated based on their specific activities. A value of 200~Bq $^{210}$Pb$\cdot$kg$^{-1}$ Pb was considered for Pb~\cite{20}.}
\label{tab:3}
\end{table}

Among the cleanroom locations, accumulation rates  in locations B, G and J were measured at 10$^{-4}$~$\mu$Bq$\cdot$day$^{-1}\cdot$cm$^{-2}$ for $^{40}$K and 10$^{-6}$~$\mu$Bq$\cdot$day$^{-1}\cdot$cm$^{-2}$  for $^{232}$Th and $^{238}$U, about one order of magnitude higher compared to those measured in the other cleanroom locations (A, C, D, E, I and K) for the three radionuclides. In location E, only accumulation rate for $^{238}$U matched those observed in B, G and J, while $^{40}$K and $^{232}$Th depositions are comparable to all the other cleanroom locations. Differences in measured accumulation rates of contaminants reflect the location of the cleanroom in the underground facility, local activities and activities carried out in adjacent areas. An event of dust rate saturation was registered by the dust monitor in location B on one day during the vial exposure. Location G is situated between two non-2,000-class-cleanroom areas: the already mentioned transition area between the mine and the clean lab (location F, or dirty carwash) and a refuge room. Moreover, location G is the main entrance for all materials introduced in the laboratory. After 20 days of reduced activity at SNOLAB during the Creighton Mine shutdown, operations resumed at SNOLAB for the rest of the vial exposure. Drilling work on the wall opposite to the vials was performed in location J during a week of the exposure time; it might have affected the results from this location.  Excluding locations B, G and J for the aforementioned reasons, average accumulation rates and their standard deviations were calculated for $^{40}$K, $^{232}$Th and $^{238}$U as (5.7$\pm$5.9)$\cdot$10$^{-5}$, (6.3$\pm$7.8)$\cdot$10$^{-7}$ and (6.5$\pm$4.4)$\cdot$10$^{-7}$~$\mu$Bq$\cdot$day$^{-1}\cdot$cm$^{-2}$, respectively. A spread of about one order of magnitude in deposition rates is noticed for $^{40}$K, $^{232}$Th and $^{238}$U and three orders of magnitude for $^{210}$Pb. Levels measured in these locations, as listed in Table~\ref{tab:3}, are compatible with those obtained in the cleanroom at PNNL which was measured to be better than class 1,000. Indeed, the monitoring of particulate counts via the laser dust monitor at SNOLAB showed counting rates compatible with a cleanroom class better than 1,000 during the exposure time. Highest levels of $^{40}$K, $^{232}$Th and $^{238}$U accumulation rates were measured in the two non-cleanroom locations, F and H.   
    ~ 

The wide variability of $^{210}$Pb accumulation rates observed among cleanroom locations reflects the lead material storage/handling history underground at SNOLAB. The highest rate was measured in location A, where, during the exposure period, bricks of lead were stored and handled for the construction of a lead shielding of a HPGe detector. The overall spread in cleanroom locations spans over four orders of magnitudes, 10$^{-10}$ to 10$^{-7}$~$\mu$Bq$\cdot$day$^{-1}\cdot$cm$^{-2}$.
At non-cleanroom locations, values were recorded at 10$^{-8}$ and 10$^{-7}$~$\mu$Bq$\cdot$day$^{-1}\cdot$cm$^{-2}$. Among the four investigated radioisotopes, $^{40}$K contributed the most radioactivity from dust, as also observed at PNNL (Figure~\ref{fig:2}). 

While the direct detection data employing the ICP-MS method described above showcases some baseline accumulation rates for PNNL and SNOLAB during moments of normal (PNNL) and minimal (SNOLAB) activity, our dust investigation method could be employed during a wide variety of activities and in different locations. For example, rotating exposures of witness vials could be employed to understand accumulation rates during specific moments of installation, setup, existence, etc. Depending on the sensitivity required, exposure times could range from a single day to many hundreds of days. Rotating exposure vial investigations could be setup to disentangle contamination steps during complex, multistep installation processes, or understand contributions from dust during moments of high traffic or unusual activity. Moreover, the same method outlined here for the analysis for K, Pb, Th, and U, could also be employed to assess contributions from nearly the entire periodic table. Such an approach may help ascertain the major source contributors to the dust (e.g. soil, concrete, machinery, etc.), or understand impurities that enter processes that may compromise performance (e.g., Ba-tagging applications).    
Given the accumulation rates above from an unoccupied SNOLAB environment, let us assume we are installing a large detector component, such as the 12~m diameter acrylic vessel from the SNO+ detector~\cite{25}, in one of the monitored locations, for example location C. The intrinsic $^{232}$Th and $^{238}$U in the acrylic has been measured in the $\mu$Bq/kg or lower~\cite{23} for each isotope, resulting in a total activity of the order of the order of 10$^{2}$~$\mu$Bq from $^{232}$Th and $^{238}$U for the total mass of the vessel ($\approx$30~ton). The radioactive contamination accumulated on the entire surface of such a component exposed to dust in location C can be calculated from data in Table~\ref{tab:3} and Figure~\ref{fig:5}, based on the vessel dimensions. For a 12~m diameter sphere in location C, accumulated radioactive contamination would be about 2~mBq$\cdot$month$^{-1}$ from $^{40}$K, 3$\cdot$10$^{-2}$~$\mu$Bq·month$^{-1}$ from $^{210}$Pb, 2$\cdot$10$^{1}$~$\mu$Bq$\cdot$month$^{-1}$ from $^{232}$Th and 5$\cdot$10$^{1}$~$\mu$Bq$\cdot$month$^{-1}$ from $^{238}$U. While these activities seem  small and not significant compared to the activity from the vast amount of material itself, it is  very much useful to have such data to validate these assumptions and that no unusual moments of ``high activity'' is happening. Moreover, it should be noted that these values correlate to accumulation rates taken in SNOLAB during moments of little-to-no activity. In future studies, we will assess dust accumulation rates in SNOLAB during time periods of ``normal'' activity. 

    
A typically assumed conservative dust fallout rate is of 10 ng $\cdot$hr$^{-1}\cdot$cm$^{-2}$ in a class 1,000 cleanroom. A fallout rate of 1-7~ng$\cdot$hr$^{-1}\cdot$cm$^{-2}$ was measured in 1995 at SNOLAB, when the laboratory was a class 2,500 cleanroom~\cite{26}. 
Based on the relative concentrations of K, Th and U in the soil provided by the USGS and the accumulation rates measured for the three elements (Table~\ref{tab:2}), dust accumulation rates of  (3.3$\pm$0.5)$\cdot$10$^{-3}$, (3.3$\pm$0.5)$\cdot$10$^{-3}$, (3.3$\pm$0.5)$\cdot$10$^{-3}$~ng$\cdot$hr$^{-1}\cdot$cm$^{-2}$ are obtained respectively from K, Th and U in the cleanroom at PNNL (nominally class 10,000, measured better than class 1,000), with a total average of (1.3$\pm$0.9)$\cdot$10$^{-3}$~ng$\cdot$hr$^{-1}\cdot$cm$^{-2}$, four orders of magnitude lower than the assumed one. Average dust accumulation rates in the cleanroom locations investigated at SNOLAB are calculated, in ng$\cdot$hr-1$\cdot$cm$^{-2}$, as (1.4$\pm$1.7)$\cdot$10$^{-2}$ from K, (1.1$\pm$2.1) from Pb, (7$\pm$15)$\cdot$10$^{-3}$ from Th and  (9$\pm$16)$\cdot$10$^{-3}$ from U. 
A wide spread of values is observed, due to the spread of accumulation rate data obtained from the various cleanroom locations. A total average dust accumulation rate from all elements in all locations of the class 2,000 cleanroom at SNOLAB is (2.7$\pm$2.6)$\cdot$10$^{-1}$~ng$\cdot$hr$^{-1}\cdot$cm$^{-2}$ and (1.0$\pm$0.4)$\cdot$10$^{-1}$~ng$\cdot$hr$^{-1}\cdot$cm$^{-2}$ if excluding data from Pb (showing a very wide spread). Final averages for the class 2,000 cleanroom, including and excluding Pb data, are, respectively one and two orders of magnitude lower compared to the assumed deposition rate of 10~ng of dust$\cdot$hr$^{-1}\cdot$cm$^{-2}$ in a class 1,000 cleanroom. 

\subsection{Comparison with estimated data }
A system of witness plates is in use at SNOLAB to monitor dust particulate fallout in various locations of the underground facility. The exposed plates are analyzed using X-Ray Fluorescence (XRF) for mine dust and surrogate elements from the shotcrete, Fe and Ca respectively, to give an idea of the radiocontaminant contribution from dust. Dust fallout rates are inferred from determinations of Ca and Fe over time, assuming the dust is from mine dust and/or shotcrete. A description of how Fe and Ca XRF analyses are performed is reported in~\cite{23}. Briefly, based on the analysis of rock samples and shotcrete samples from the underground laboratory previously performed~\cite{27}, the relative abundance of Th and U to (the detectable) Ca and Fe is known. Assuming collected dust is composed of mainly mine dust and/or shotcrete, the radionuclide fallout rate is extrapolated from the measured Ca and Fe (which are at much higher concentrations in the dust than Th and U) fallout rates. 
\begin{table}
\resizebox{\textwidth}{!}{
\begin{tabular}{|c|c|c|c|c|}
\hline 
 \multirow{2}{*}{Location } &	\multicolumn{2}{|c|}{Fe [ng$\cdot$day$^{-1}\cdot$cm$^{-2}$]} & \multicolumn{2}{|c|}{Ca [ng$\cdot$day$^{-1}\cdot$cm$^{-2}$]}\\
 & 	XRF & ICP-MS & XRF & ICP-MS \\
\hline
B 
&
0.10 $\pm$ 0.07 
&
0.035 $\pm$ 0.008 
&
$<$1.8 
&
0.05 $\pm$ 0.03 \\
C 
&
$<$0.13 
&
0.1 $\pm$ 0.2 
&
$<$1.8 
&
0.09 $\pm$ 0.04 \\
E 
&
0.33 $\pm$ 0.08 
&
0.015 $\pm$ 0.007 
&
$<$1.8 
&
0.03 $\pm$ 0.01 \\
F 
&
4.0 $\pm$ 0.2 
&
4.2$\pm$ 0.9 
&
4.30 $\pm$ 0.07 
&
1.7 $\pm$ 0.2 \\
G 
&
$<$0.13 
&
0.07 $\pm$ 0.07 
&
$<$1.8 
&
0.08 $\pm$ 0.04 \\
I 
&
0.23 $\pm$ 0.07 
&
0.04 $\pm$ 0.05 
&
$<$1.8 
&
0.02 $\pm$ 0.01 \\
\hline
\end{tabular}}
\caption{Accumulation rates of Fe and Ca, reported in ng per day per square cm of surface area, measured via XRF at SNOLAB on the witness plates and via ICP-MS at PNNL on the PFA vials over the same period of exposure.  Averages and standard deviations were calculated from replicates (n=4).}
\label{tab:4}
\end{table}

\begin{figure}[hbt!]
     \begin{minipage}{0.35\textwidth}
        \centering
        \includegraphics[scale=0.35]{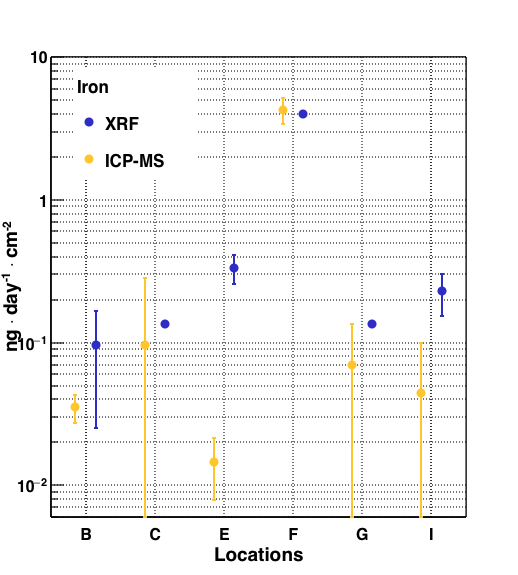}
    \end{minipage}\hfill
    ~ 
  \begin{minipage}{0.35\textwidth}
        \centering
        \includegraphics[scale=0.35]{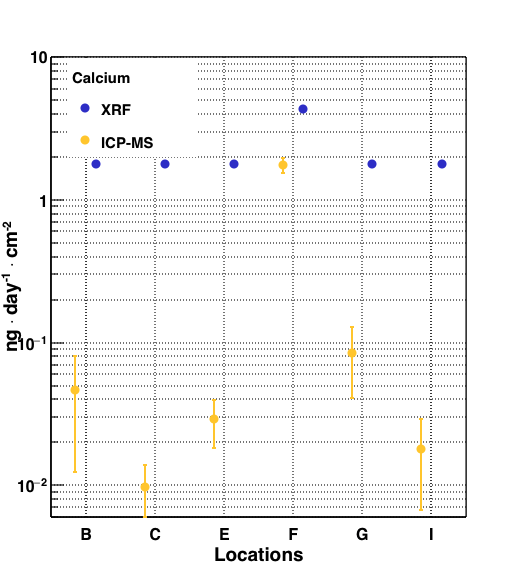}
   \end{minipage}\hfill
    ~ 
    \caption{Comparison between accumulation rates measured via XRF at SNOLAB on the witness plates and via ICP-MS at PNNL on the PFA vials in locations B, C, E, F, G and I over the same exposure period. All accumulation rates are reported in ng per day per square cm. Left panel: accumulation rates for iron. Right panel: accumulation rates for calcium. }\label{fig:6}
\end{figure}

The dust collection experiment in this work, when possible, has targeted locations at SNOLAB also monitored by the witness plate method. In particular, witness plates are located in locations B, C, E, F, G and I. Data for Ca and Fe accumulation rates from XRF analysis of witness plates relative to the PFA vial exposure period were compared to ICP-MS determinations of Fe and Ca accumulation rates on the PFA vials.  Table~\ref{tab:4} reports data obtained from the two techniques over the same exposure time frame. Comparisons between XRF and ICP-MS data are also shown in Figure~\ref{fig:6} for Fe (left panel) and Ca (right panel).  Results are reported in ng of analyte deposited per day per square cm of exposed surface. Results from ICP-MS analyses are reported as the average value and standard deviation of four independent replicates. XRF data are reported as the results single replicates; the uncertainty associated to each value was estimated based on counting statistics. DLs for XRF measurements were calculated as 3$\cdot$StdDev of a process blank (unexposed witness plate). XRF results for Fe were all above detection limit, 0.13~ng$\cdot$day$^{-1}\cdot$cm$^{-2}$, except for locations C and G. 

Values for XRF analysis in these locations in the left panel of Figure~\ref{fig:6} represent therefore upper limits. For Ca, instead, XRF measurements were all below detection limit, 1.8~ng$\cdot$day$^{-1}\cdot$cm$^{-2}$, with the exception of the transition area between the mine and the clean laboratory (dirty carwash, location F). All values for XRF measurements in the right panel of Figure~\ref{fig:6} are upper limits; only the result reported for location F is a measured value.    

 For Fe, XRF and ICP-MS data are overall compatible in all locations, except for a discrepancy of about one order of magnitude in locations E.  In this location, the vials were not adjacent to the witness plate. They might have been exposed to a different air flow causing the observed inconsistency. A large variability among replicates of the same set of vials was observed in ICP-MS results for Fe in locations C, G and I, which resulted in high standard deviations for these values. ICP-MS data for Ca are compatible with detection limits from XRF data. A small discrepancy (factor of $\approx$1.5) was observed between XRF and ICP-MS results in location F. The comparison between analyses for Fe and Ca via XRF and ICP-MS over the same exposure period and their compatibility validates both techniques for the measurements of Ca and Fe fallout rates from dust. However, ICP-MS sensitivities allowed for quantitation of both elements in all locations. XRF sensitivity was not sufficient to quantify Ca on witness plates after 42-day exposure in cleanroom locations, where only upper limits were obtained. Quantitation of Ca via XRF was instead possible in the non-cleanroom location F (dirty carwash). Upper limits from XRF analysis were also obtained for Fe in locations C and G, where ICP-MS still was able to provide actual values. 

\begin{table}
\centering
\begin{tabular}{| c | c | c |}
\hline 
Element & Rock & Concrete\\
\hline
K [\%]  & 0.99 & 1.6 \\
Ca [\%] & 3.6 & 10.1\\  
Fe [\%] & 6.5 & 2.6 \\
Pb [ppm] & 10.4 & 13.9 \\
Th [ppm] & 5.4 & 13.1  \\
U [ppm] & 1.2 & 2.4 \\
\hline
\end{tabular}
\caption{Concentrations of K, Ca, Fe, Pb, Th and U in rock and concrete at the SNOLAB site~\cite{27}.}
\label{tab:5}
\end{table}

 Based on the composition of rock and concrete at the SNOLAB site reported in~\cite{27}, fallout rates for K, Pb, Th and U were extrapolated from Fe and Ca measured fallout rates in locations B, C, E, F, G and I. Table~\ref{tab:5} lists concentration values of the considered elements in the rock (average values for samples analyzed in~\cite{27}) and in the concrete. 
 
 The relative concentrations of Fe, Ca, K, Pb, Th and U in the rock and concrete were used to estimate fallout rates, using the following equation:  
\begin{equation}
F_{e_{X}} = F_{m_{Y}} \cdot \frac{C_{X}}{C_{Y}}  
\end{equation} 

\begin{figure}[htb!]
    \begin{subfigure}
      \centering
      \includegraphics[width=1.1\textwidth]{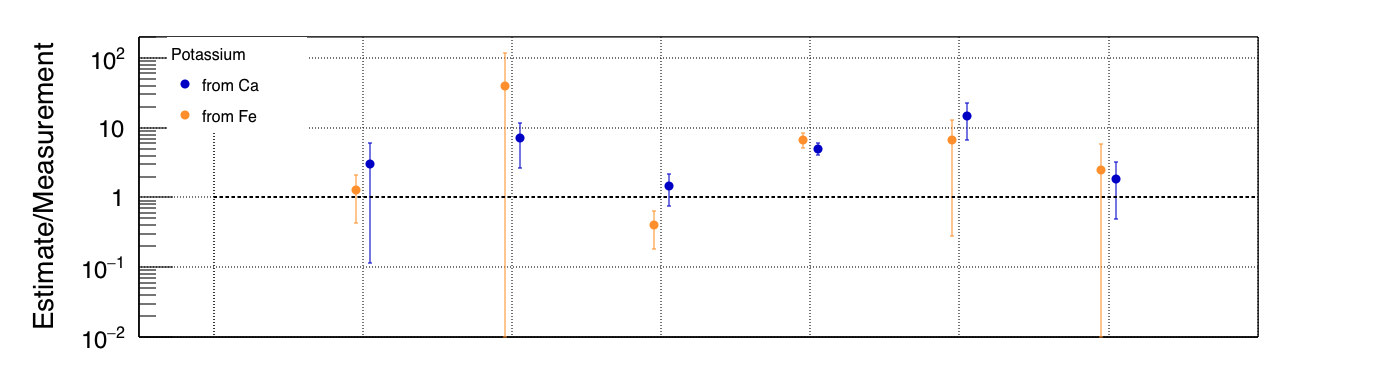}
    \end{subfigure}
        \begin{subfigure}
      \centering
      \includegraphics[width=1.1\textwidth]{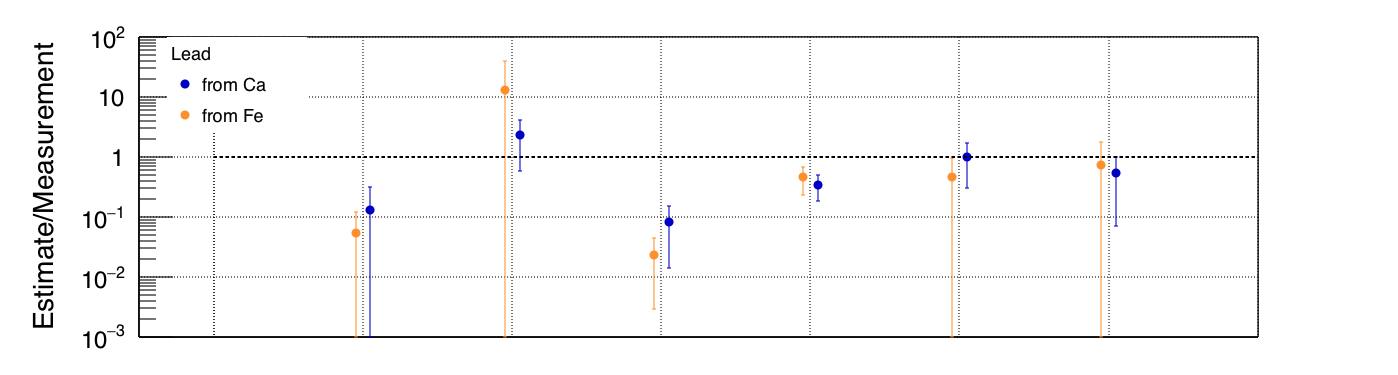}
    \end{subfigure}
        \begin{subfigure}
      \centering
      \includegraphics[width=1.1\textwidth]{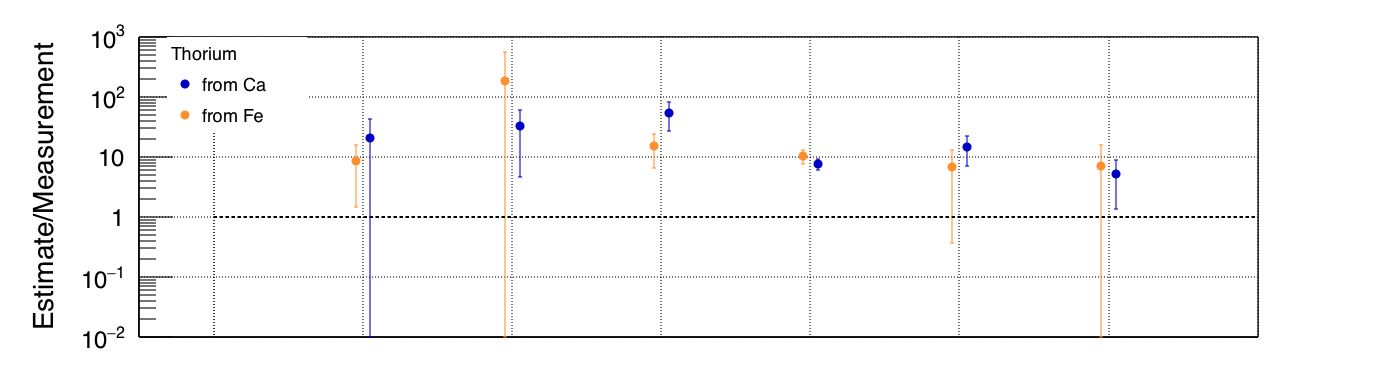}
    \end{subfigure}
        \begin{subfigure}
      \centering
      \includegraphics[width=1.1\textwidth]{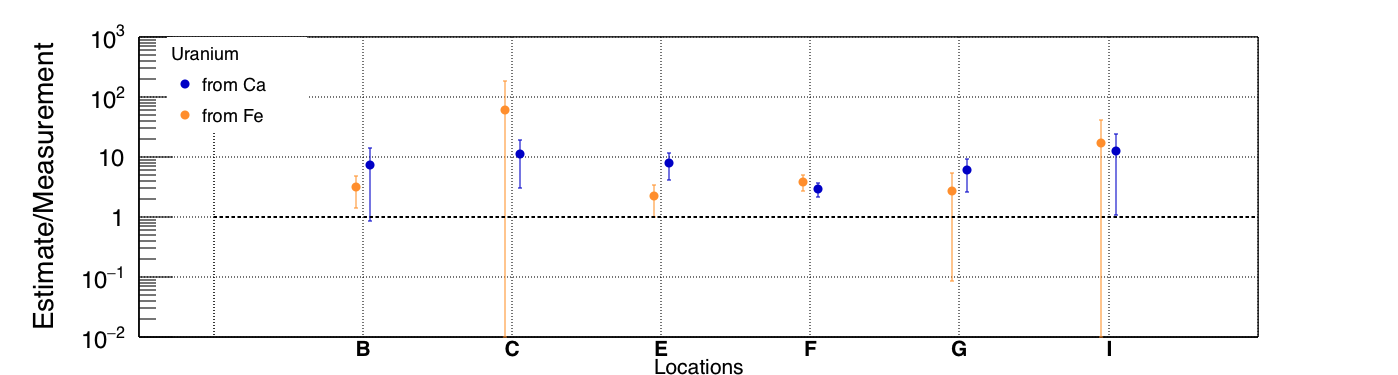}
    \end{subfigure}
 \caption{From top to bottom: ratios of estimated over ICP-MS measured fallout rates for K, Pb, Th and U for  locations B, C, E, F, G and I, assuming the mine dust being the only source of particulate contamination. Concentrations of elements in the rock at the SNOLAB site (Table~\ref{tab:3}) were used to estimate fallout rates for each element both from Ca (blue markers) and from Fe (orange markers).}
    \label{fig:7}
\end{figure}

Where F$_{e_{X}}$ is the estimated fallout rate for element X (K, Pb, Th or U), F$_{m_{Y}}$ is the measured fallout rate of element Y (Fe or Ca) and C$_{X}$ and C$_{Y}$ are the concentrations of elements X and Y in the rock or concrete. Estimated fallout rates were compared to fallout rates directly measured via ICP-MS. Estimates were obtained for each of the four investigated elements (K, Pb, Th and U) from measured fallout rates of both surrogate elements Fe and Ca. Moreover, calculations were performed in both assumptions of mine dust (rock) and concrete being the sole source of dust particulate. For Fe and Ca, data obtained from the ICP-MS analysis were used, given that ICP-MS provided actual values, not upper limits, for both elements in all considered locations. A comparison between estimated and measured data for K, Pb, Th and U was performed by plotting ratios of estimated over measured fallout rate for each element. Figure~\ref{fig:7} shows the ratios for all elements, in the assumption of mine dust as the only source of particulate. For each element both data obtained from Ca measurements (blue markers) and Fe measurements (orange markers) are plotted. A black dotted line is set at 1 for the ideal case in which the estimated and the measured fallout rates match.    

\begin{figure}[htb!]
    \begin{subfigure}
      \centering
      \includegraphics[width=1.1\textwidth]{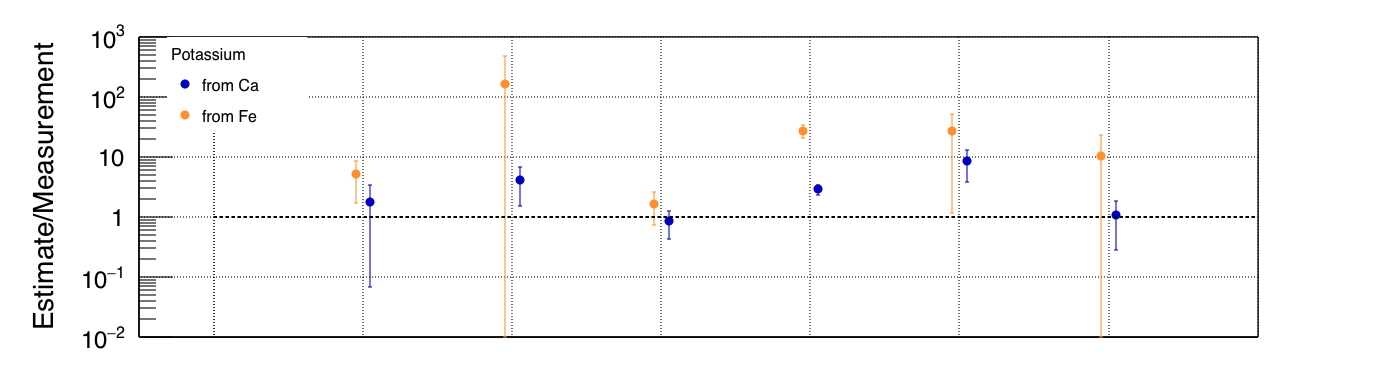}
    \end{subfigure}
        \begin{subfigure}
      \centering
      \includegraphics[width=1.1\textwidth]{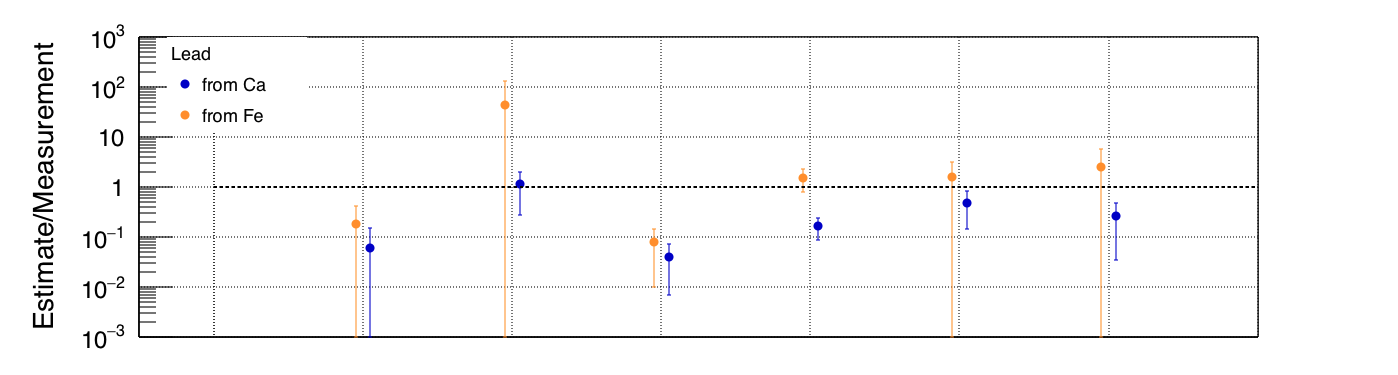}
    \end{subfigure}
        \begin{subfigure}
      \centering
      \includegraphics[width=1.1\textwidth]{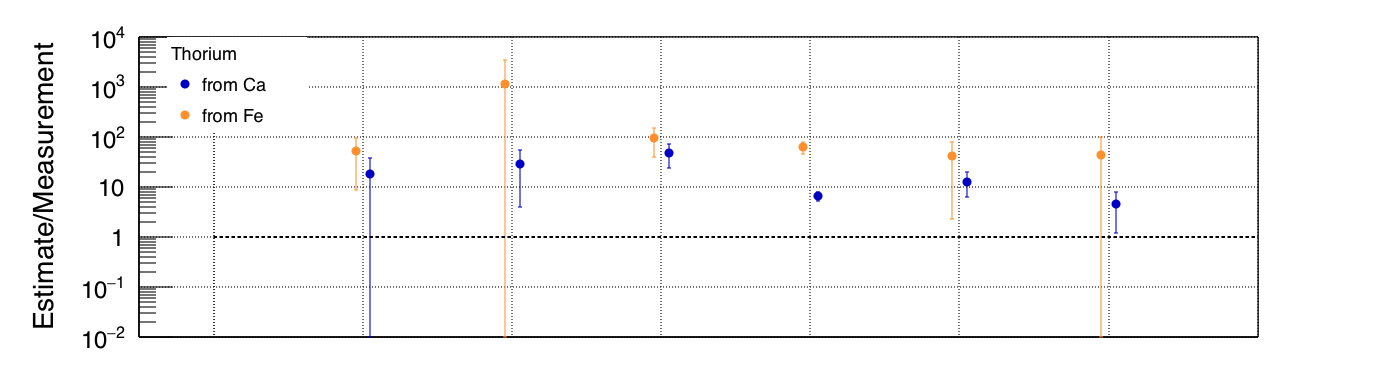}
    \end{subfigure}
        \begin{subfigure}
      \centering
      \includegraphics[width=1.1\textwidth]{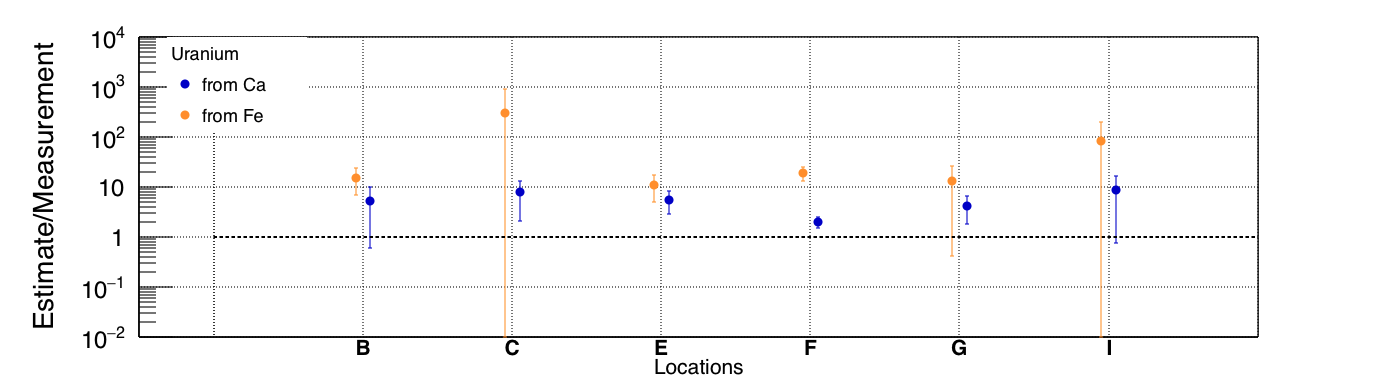}
    \end{subfigure}
 \caption{From top to bottom: ratios of estimated over ICP-MS measured fallout rates for K, Pb, Th and U for  locations B, C, E, F, G and I, assuming the concrete being the only source of dust particulate. Concentrations of elements in the rock at the SNOLAB site (Table~\ref{tab:3}) were used to estimate fallout rates for each element both from Ca (blue markers) and from Fe (orange markers).}
    \label{fig:8}
\end{figure}

Estimates of fallout rates for K are compatible with measured values in locations B and I. Values estimated from Ca are compatible with measured values in location E, where estimates of rates for K from Fe provides instead values lower than those directly measured. In location G, data estimated from Fe are compatible with measured data due to the relatively high standard deviation. In all the other locations, data inferred from both from Fe and Ca overestimate K fallout rates of about one order of magnitude compared to directly measured values. With the exception of location C, where estimates provide fallout rates higher (up to one order of magnitude for data from Fe) than those measured, estimates for Pb provide numbers lower or compatible to those measured. In particular, in location E inferring data from Ca underestimates Pb fallout rate of one order of magnitudes and two orders of magnitude if using Fe as the surrogate element. The difference between estimated and measured data for Pb in location F is within one order of magnitude. Inferred and measured data for Pb are compatible in locations G and I. Inferring data for Th from Fe and Ca measurements overestimated Th accumulation rates of about one order of magnitude in all investigated locations. Only in location G, data estimated from Fe are compatible with measured value due to a large standard deviation. In location C and E data estimated for Ca were almost two orders of magnitudes above the measured ones and in location C data estimated from Fe were more than two orders of magnitude higher compared to direct measurements. The same trend described for Th data was observed for U, as visible in the bottom plot of Figure~\ref{fig:7}. 
Data obtained assuming concrete as the only source of dust particulate contamination - using Ca and Fe determinations to estimate K, Pb, Th and U fallout at SNOLAB - are reported in Figure~\ref{fig:8}. As in Figure~\ref{fig:7}, each graph reports ratios of estimated over directly measured fallout rates. Data for K, Pb, Th and U are shown in Figure~\ref{fig:8}, from the top to the bottom panel, respectively. In each panel, orange and blue data points refer to data obtained estimating fallout rates from measured Fe and Ca rates respectively. The same trends described for data obtained with the assumption of all particulate coming from mine dust (Figure~\ref{fig:7}) are observed. Data for K estimated from Ca are compatible with directly measured data in locations B, E and I; a discrepancy of a factor of about 5 is observed in locations C, F and G.  Data for K estimated from Ca are compatible with directly measured data in locations B, E and I; a discrepancy of a factor of about 5 is observed in locations C, F and G. Data for K inferred from Fe measurements overestimate K fallout rates in all locations by one to two orders of magnitudes in all locations except location E, where the estimated value is close to the measured one. Inferring data for Pb from Ca underestimated fallout rates in all locations, except C, where estimated and measured fallout rates are comparable. Inferring data for Pb from Fe still underestimated Pb fallout rates in locations B and E, data were comparable to measured values in locations F, G and I, while an overestimate of almost two orders of magnitude was observed in location C. Inferring data for Th from both Ca and Fe provided accumulation rate values higher of at least one order of magnitude compared to what was obtained from a direct ICP-MS measurement. A discrepancy of about one order of magnitude between estimated and measured data is observed for U. Data estimated from Fe in location C and data estimated from Ca in locations G and I are compatible with measured data due to the large standard deviation of the data.  

Comparisons between measured and estimated data demonstrated that inferring contaminant fallout rates from predictor or surrogate elements, based on assumed dust composition, is a viable method to monitor anomalous variations in particulate fallout and obtain rough estimates. However, the composition of dust in controlled environments (e.g., cleanrooms) where airflows are filtered, does not necessarily reflect the local composition of soil. Dust in cleanrooms is mainly generated by the grinding of handled materials, cleanroom garbs and other human activities. Particulate fallout rate and its chemical composition in such environments, therefore, are not to be considered constant and accurately predictable based on models and assumed compositions. Variations occur depending on the ongoing activities. The lead fallout measured in location A, where lead was stored and handled during the exposure time, compared to the other cleanrooms (Figure \ref{fig:5}, Table \ref{tab:3}) is a demonstration. Inferring fallout rates of contaminants from dust using predictors and models can result in inaccurate estimates, either overestimated or underestimated. In this study, discrepancies of one to two orders of magnitude have been observed. 

\section{Discussions and Conclusion}
This work provides an effective and valuable method for the direct measurement of contaminant accumulation rates on material surfaces after exposure to dust. Dust collection is made extremely practical by the use of ultralow background PFA vials, already used in a variety of ultrasensitive analyses~\cite{12} and \cite{28}-\cite{31}, as collection media. Exposure can be easily performed in any location of interest, only requiring an operator available for vial exposure and recapping. The possibility to recap collection media after exposure allows for transportation of the media to facilities equipped with an ultrasensitive ICP-MS laboratory, when such a facility is not available on site, without any loss of information. Any research or production facility where particulate contamination from dust is critical to activities can significantly benefit from the method we have developed and proposed in this work. Given that the ICP-MS technique can reach unique sensitivities when coupled with ultraclean procedures, as also demonstrated by the comparison between data obtained at SNOLAB by XRF and ICP-MS over the same exposure time, reasonable exposure times (from one day to about one month, depending on the environment classification) are required to investigate dust fallout even in cleanrooms, without the need of any expensive dust collection equipment.
While originally developed to quantify long-lived radionuclides and stable Pb, the method can potentially investigate any stable element in the periodic table, providing a very localized elemental fingerprint to dust sources that could be utilized to obtain insight into background sources of dust. Exploiting the ICP-MS multi-element capabilities, target elements can be directly monitored, eliminating the need to infer data based on models and/or assumptions, which, as we have demonstrated, can result in significant inaccuracies. Dust fallout rates and composition in cleanrooms strongly depend on the local ongoing activities. Data reported in this work refer to a period of regular activity at PNNL and reduced activity at SNOLAB. New campaigns of measurement at SNOLAB are planned and some of them have already started. We intend to target a larger number of elements in the future campaigns, aiming at locally identifying the major contributors to dust. Moreover, we intend to utilize the method to quantitatively test dust removing procedures ($e.g.,$ blowing with pure nitrogen, wiping or spraying material surfaces). Although developed to benefit the design and construction of next generation ultrasensitive detectors studying rare events in the field of fundamental physics, the method proposed in this work can be advantageously exploited in all research or industry applications where ultralow levels of contamination, stable or radioactive, from dust particulate is critical and needs to be monitored and/or rejected \cite{33}\cite{34}.  

\section{Acknowledgments}
Pacific Northwest National Laboratory (PNNL) is operated by Battelle for the United States Department of Energy (DOE) under Contract no. DE-AC05-76RL01830. This study was supported by the DOE Office of High Energy Physics Advanced Technology R\&D subprogram.
The authors would like to thank SNOLAB and its staff for support through underground space, logistical and technical services. SNOLAB operations are supported by the Canada Foundation for Innovation and the Province of Ontario Ministry of Research and Innovation, with underground access provided by Vale at the Creighton mine site.

\end{document}